\documentclass[twocolumn]{svjour3} 

\usepackage{graphicx}
\usepackage{amsmath}
\usepackage{color}
\usepackage{array}
\usepackage{color}

\renewcommand\k{\mathbf{k}} 

\newcommand{\loc}{\mathrm{loc}}

\journalname{Journal of Superconductivity and Novel Magnetism}

\begin{document}

\title{Avoiding Stripe Order: Emergence of the Supercooled Electron Liquid}

\author{Louk Rademaker \and
	Arnaud Ralko \and
	Simone Fratini \and
	Vladimir Dobrosavljevi\'{c}
} 
\institute{L. Rademaker \at
	Kavli Institute for Theoretical Physics, University of California Santa Barbara, CA 93106, USA \\
	\email{louk.rademaker@gmail.com}
\and
	A. Ralko \at
	Institut N\'{e}el-CNRS and Universit\'{e} Grenoble Alpes, F-38042 Grenoble Cedex 9, France \\
\and
	S. Fratini \at
	Institut N\'{e}el-CNRS and Universit\'{e} Grenoble Alpes, F-38042 Grenoble Cedex 9, France \\
\and
	V. Dobrosavljevi\'{c} \at
	Department of Physics and National High Magnetic Field Laboratory, Florida State University, Tallahassee, Florida 32306, USA
	}

\date{\today}


\maketitle

\begin{abstract} 
In the absence of disorder, electrons can display glassy behavior through supercooling the liquid state, avoiding the solidification into a charge ordered state. Such supercooled electron liquids are experimentally found in organic $\theta$-$MM'$ compounds. We present theoretical results that qualitatively capture the experimental findings. At intermediate temperatures, the conducting state crosses over into a weakly insulating pseudogap phase. The stripe order phase transition is first order, so that the liquid phase is metastable below $T_s$. In the supercooled liquid phase the resistivity increases further and the density of states at the Fermi level is suppressed, indicating kinetic arrest and the formation of a glassy state. Our results are obtained using classical Extended Dynamical Mean Field Theory.
\end{abstract}

\section{Introduction}

Glassy dynamics of electrons in the absence of disorder\cite{Debenedetti2001,Schmalian:2000fn} has long been elusive. Instead, quenched disorder seemed necessary for the kinetic arrest of electrons\cite{Pollak:2013vs}. However, recent experiments on $\theta$-$MM'$ organic compounds\cite{Kagawa2013,monceau2007prb,Sato2014PRB} found an exponential slowing down of the electron dynamics in a disorder-free environment. Inspired by these results, we proposed a model where the geometric frustration of the triangular lattice, augmented with long-range interactions, causes this glassy behavior. Using classical Monte Carlo simulations we showed that this model indeed reproduces the Arrhenius $\tau \sim e^{\Delta/T}$ dynamics.\cite{Mahmoudian:2014vh}

So how can glassy behavior arise without quenched disorder? The answer to this question can be found by considering ordinary structural glasses, like window glass.\cite{Debenedetti2001} A first order solidification transition is avoided by fast-cooling, leading to a supercooled liquid regime. Lowering the temperature of the supercooled liquid leads to the kinetic arrest and self-generated disorder.

Similarly, in $\theta$-type organics there is a solid phase of the electrons: the ground state is known to have stripe order.\cite{Kagawa2013} The frustration caused by the triangular lattice and long-range interactions drive this transition first order, making it possible to form a supercooled electron liquid. Indeed, the cooling rate in $\theta$-RbZn determines whether it enters the stripe ordered phase or the supercooled 'glassy' phase.

In these Proceedings we present theoretical results on the density of states and the resistivity of long-range interacting electrons on a triangular lattice, in the high-temperature liquid phase, the stripe solid phase and the supercooled liquid phase. We employ the methods of Extended Dynamical Mean Field Theory (EDMFT)\cite{Georges:1996un,Pramudya:2011iv,Pankov:2002gh,Pankov:2005fe,Muller:2007hya}. Our results qualitatively capture the physics of the supercooled electron liquid, obtained by avoiding stripe order.

\section{The model}

\subsection{Long-range interactions on a triangular lattice}
The starting point is the model introduced in Ref. \cite{Mahmoudian:2014vh}, of spinless interacting electrons on a triangular lattice. In the limit where the quantum hopping $t$ is much smaller than the interaction strength $V$, the Hamiltonian is given by the interaction terms
\begin{equation}
	H = \frac{1}{2} \sum_{ij} V_{ij} 
	\left(n_i - \frac{1}{2} \right) 
	\left(n_j - \frac{1}{2} \right).
	\label{Hamiltonian}
\end{equation}
The electron density is fixed at one electron per two lattice sites. With only nearest neighbor interactions, this model corresponds to the triangular Ising model which is known to have a macroscopic ground state degeneracy\cite{Wannier:1950}. We consider here instead the long-range Coulomb interaction between electrons, which lifts this degeneracy.
\begin{equation}
	V_{ij} = \frac{V}{|\mathbf{r}_{ij}|}
\end{equation}
where $V$ is the nearest neighbor repulsion, and $\mathbf{r}_{ij}$ is measured in units where the lattice constant is $a=1$.

The ground state of Eqn. (\ref{Hamiltonian}) with Coulomb interaction is given by the stripe ordered state, with wavevector $\mathbf{k} = M = \left( 0, \frac{2\pi}{\sqrt{3}} \right)$.\cite{Mahmoudian:2014vh} At this $M$-point, the Fourier transform of the Coulomb interaction equals
\begin{equation}
	V_{\mathbf{k} = M} = -1.40447 V,
\end{equation}
from which one can find the stripe ground state energy, $E_0 = V_M/8$. In the stripe ordered phase, the density of states consists of two delta peaks at $\pm|V_M|/2$. Standard mean field considerations suggest a second-order phase transition at $T_c = |V_M|/4$.

\subsection{EMDFT}
\begin{figure} 
  \includegraphics[width=\columnwidth]{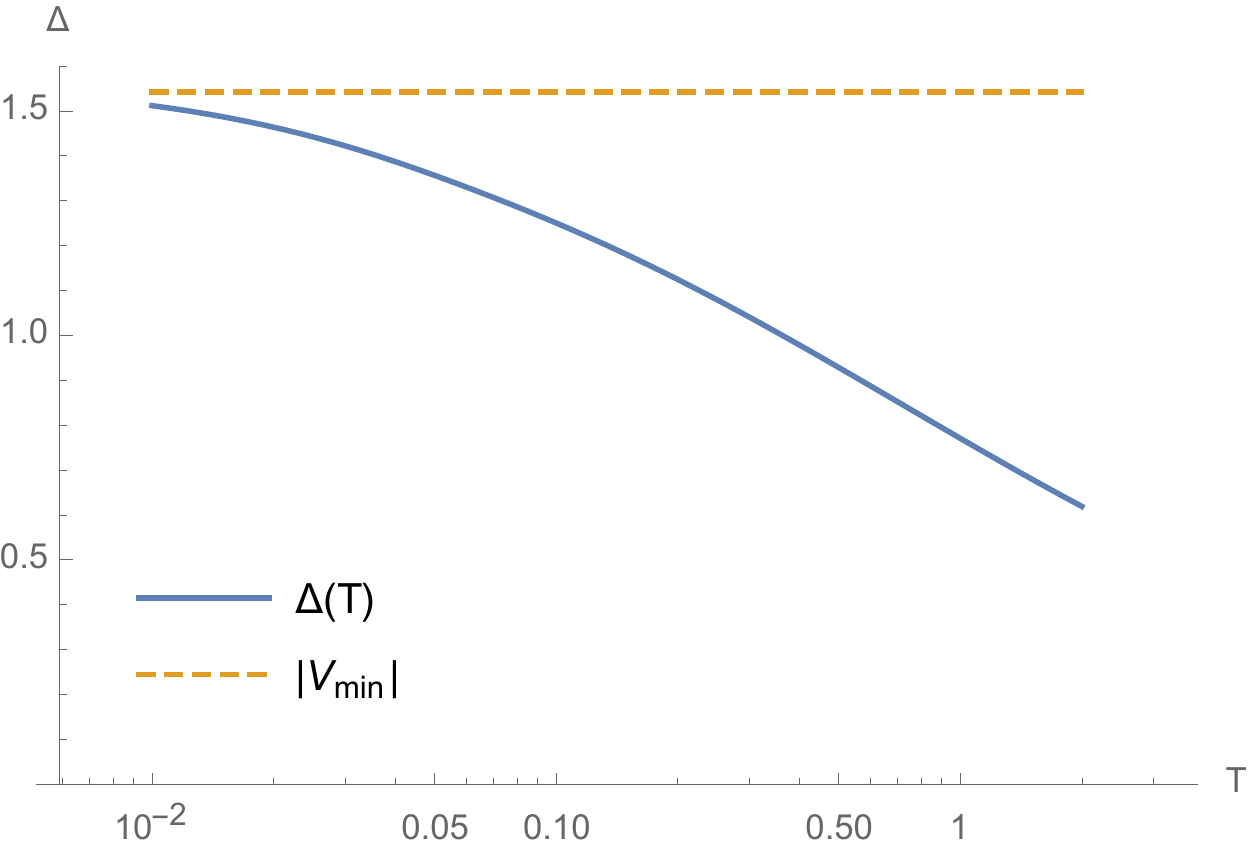}
  \caption{The self-consistent field $\Delta$ as a function of temperature $T$, as a solution of Eqn. (\ref{EDMFTeq}). The density of states and conductivity follow from this result. In the low temperature limit, $\Delta$ cannot exceed the minimum value of the interaction in momentum space, $V_{\mathrm{min}}$.}
  \label{Fig:Delta}
\end{figure}

A popular tool to study interacting electron systems is Dynamical Mean Field Theory (DMFT)\cite{Georges:1996un}, where the electron self-energy is found self-consistently assuming it is momentum-independent. Extended DMFT (EDMFT) extends this idea to the electron polarization bubble.\cite{Pramudya:2011iv,Pankov:2002gh,Pankov:2005fe,Muller:2007hya} In the classical limit $t \rightarrow 0$, where the polarization bubble becomes independent of frequency, the standard RPA expression for the polarization of an interacting system equals
\begin{eqnarray}
	\Pi_\k
		&\equiv& \sum_{\mathbf{r}_{ij}} \langle (n_i - \overline{n}) (n_j - \overline{n}) \rangle
			e^{i \k \cdot \mathbf{r}_{ij}} \nonumber \\
		&=& \frac{1}{(\overline{n} - \overline{n}^2)^{-1} + \beta ( \Delta + V_\k)}
\end{eqnarray}
with $\Delta$ the only free parameter, $\overline{n}$ the average electron density and $\beta$ the inverse temperature. For classical particles, the onsite density-density correlations are trivially $\Pi_{\loc} = \overline{n} - \overline{n}^2$. Because $\Pi_\loc = \sum_\k \Pi_\k$, we arrive at the self-consistent classical EDMFT equation
\begin{equation}
	\overline{n} - \overline{n}^2 = 
		\sum_\k \frac{1}{(\overline{n} - \overline{n}^2)^{-1} + \beta ( \Delta + V_\k)}.
	\label{EDMFTeq}
\end{equation}
In the following, we will solve this equation for $\Delta$ as a function of $T$, the results are shown in Fig. \ref{Fig:Delta}.

Many other properties of the system, within the framework of EDMFT, can be derived from $\Delta$. The local density of states $\rho (\omega)$ is given by the sum of two Gaussians,\cite{Pramudya:2011iv}
\begin{equation}
	\rho(\omega) = \frac{1}{2 \Delta} \sqrt{\frac{\beta \Delta}{2 \pi}}
		\left( e^{-\frac{\beta}{2\Delta} \left(\omega + \frac{\Delta}{2} \right)^2} 
			+ e^{-\frac{\beta}{2\Delta} \left(\omega - \frac{\Delta}{2} \right)^2}  \right).
	\label{dos}
\end{equation}
From the classical EDMFT results we can find the conductivity in a perturbative manner in leading order in the electron hopping $t \ll V$. The dc conductivity $\sigma_{DC}$ is then given by\cite{Pramudya:2011iv}
\begin{equation}
	\sigma_{DC} \sim 
		\int_{-\infty}^\infty d\omega \;
		\frac{\rho^2(\omega)}{4T \cosh^2 \frac{\omega}{2T}}
	= \frac{\beta }{8 } \sqrt{\frac{\beta }{\Delta \pi}}
		e^{-\frac{1}{4} \beta \Delta}.
	\label{sigmadc}
\end{equation}
The resulting density of states and dc conductivity are shown in Fig. \ref{Fig:DOS} and Fig. \ref{Fig:sigmaDC}, respectively.

\section{Pseudogap phase}
\begin{figure} 
  \includegraphics[width=\columnwidth]{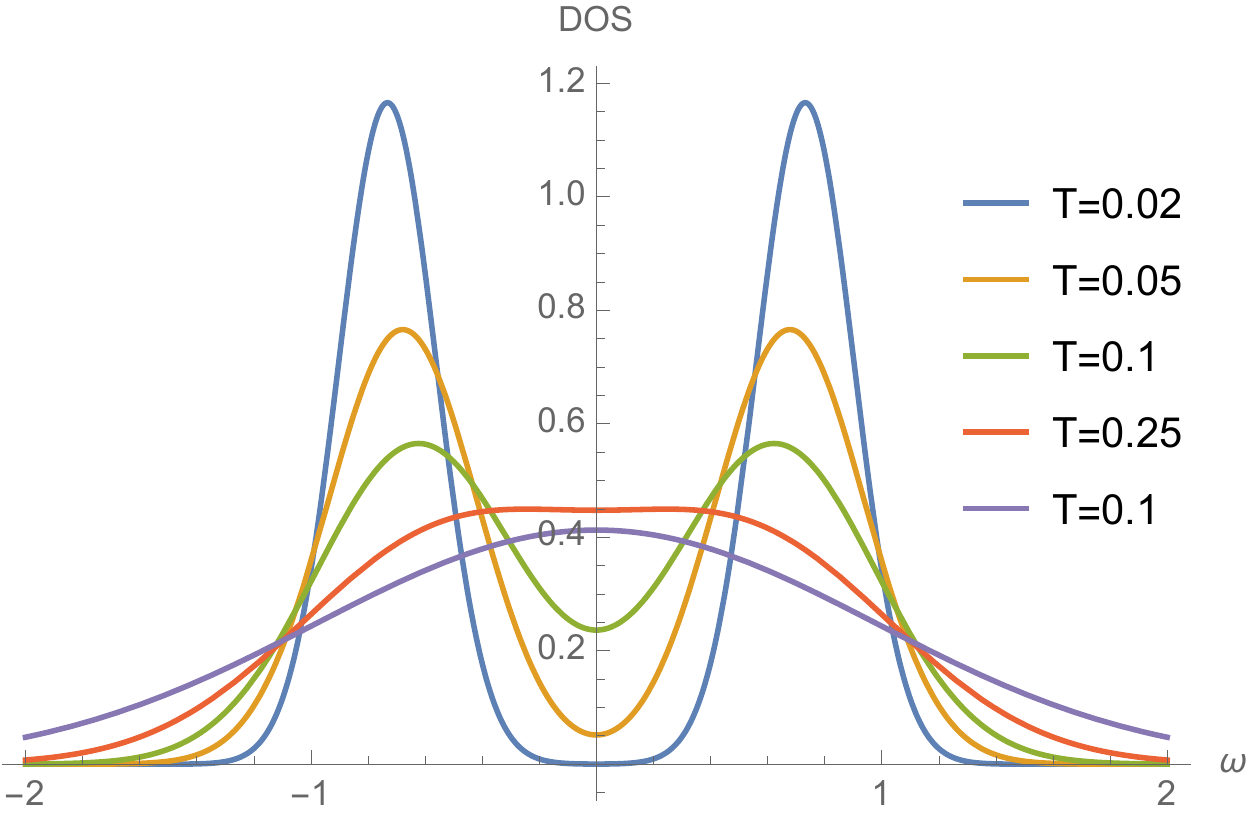}
  \caption{The local density of states following Eqn. (\ref{dos}), in the correlated liquid phase. Around $T_{PG} \approx 0.25V$ a pseudogap opens, suppressing the density of states at the Fermi level. This smoothly connects to the supercooled liquid phase where the density of states at the Fermi level is exponentially suppressed.}
  \label{Fig:DOS}
\end{figure}

\begin{figure} 
  \includegraphics[width=\columnwidth]{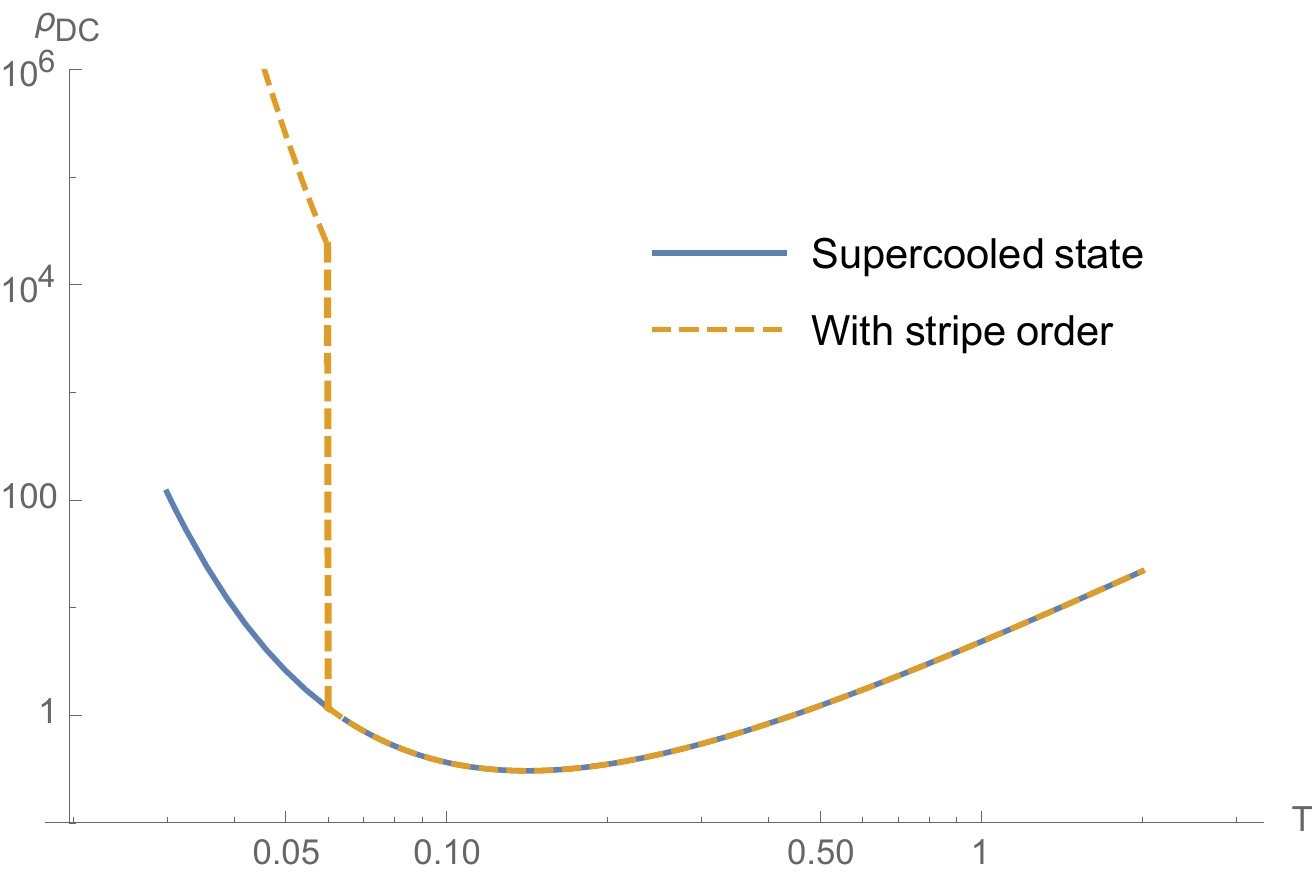}
  \caption{The dc resistivity following Eqn. (\ref{sigmadc}). At $T_{\mathrm{min}} \approx 0.14V$ the resistivity displays a minimum, indicating weakly insulating behavior and a correlated liquid phase. Below $T_s\approx 0.06V$, the stripe order sets in with a first order transition (see Fig. \ref{Fig:FirstOrder}). In the stripe ordered phase, the resistivity is activated. This result is qualitatively similar to the experimental results on $\theta$-RbZn, see Fig. 1b of Ref. \cite{Sato2014PRB}.}
  \label{Fig:sigmaDC}
\end{figure}

Given the solution of the EDMFT equations, we can analyse the different phases. At high temperatures, the system is a conducting electron liquid with resistivity increasing with temperature, as can be seen in Fig. \ref{Fig:sigmaDC}. Upon lowering the temperature, however, around $T_{PG} \approx 0.25 V$ the density of states opens a soft 'pseudogap'-like feature around the Fermi level, see Fig. \ref{Fig:DOS}. The pseudogap causes the system to become weakly insulating, with a minimum in the resistivity reached at $T_{\mathrm{min}} \approx 0.14$.

The behavior of the resistivity is qualitatively similar to experimental results in the organic compounds $\theta$-$MM'$.\cite{Kagawa2013,monceau2007prb,Sato2014PRB} Both the resistivity minimum and the pseudogap signal the crossover to a strongly correlated electron liquid. In this regime, seen in both experiments and our earlier Monte Carlo simulations, the system also develops short-range charge correlations.

\section{The stripe ordered phase}

Since the ground state is stripe ordered, one expects that at some finite temperature $T_c$ the system enters a stripe ordered phase.  To go beyond the standard mean field theory, we allow for the classical EDMFT equations to break translational symmetry. To do so, we introduce in the electron density on a given site the stripe order parameter $m$ such that
\begin{equation}
	\langle \hat{n}_i \rangle = \frac{1}{2} + m \cos \left( \mathbf{Q} \cdot \mathbf{r}_i \right) = \frac{1}{2}  \pm m.
\end{equation}
The presence of stripe order reduces the on-site density correlations,
\begin{equation}
	\Pi_{\loc,i} = \langle \hat{n}_i - \hat{n}_i^2 \rangle = \frac{1}{4} - m^2.
\end{equation}
As a result, the EDMFT equation (\ref{EDMFTeq}) is modified to
\begin{equation}
	\frac{1}{4} - m^2
		= \sum_k \frac{ 1 }
			{(\frac{1}{4} - m^2)^{-1} + \beta (\Delta +  V_k) }
\end{equation}
where the self-consistent field $\Delta$ now depends on both $m$ and temperature $T$. The stripe order parameter $m$ needs to be found from mean field theory,
\begin{equation}
	m = - \frac{1}{2} 
		\tanh \left( \frac{1}{2} m \beta (V_\mathbf{Q} + \Delta(m,T) ) \right)
	\label{MFeqnu}
\end{equation}
which can be derived from the number density expectation value. Thus for a given temperature the EDMFT equation gives $\Delta$ as a function of $m$. Then, using the mean field condition Eqn. (\ref{MFeqnu}) we find the corresponding value of $m$.

\begin{figure} 
  \includegraphics[width=\columnwidth]{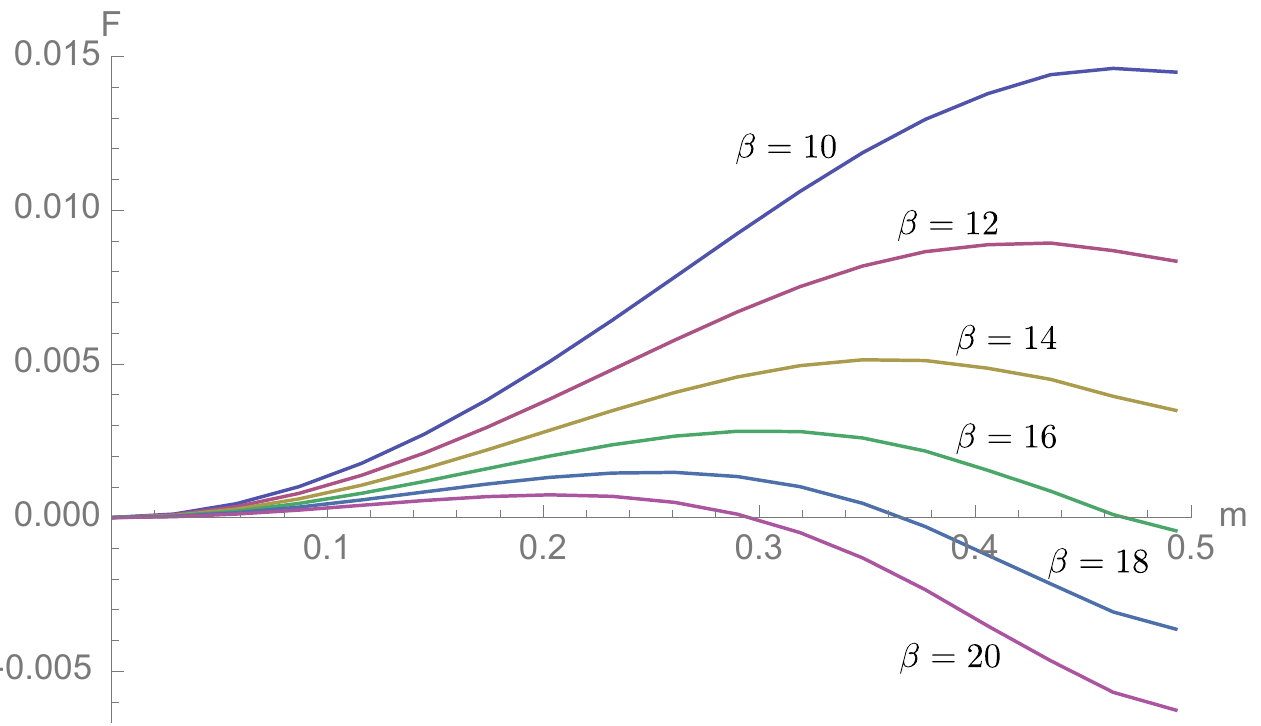}
  \caption{The free energy as a function of stripe order parameter $m$ at a given inverse temperature $\beta = 1/T $ (at values $10$, $12$, $14$, $16$, $18$ and $20$), following non-uniform EDMFT. From this we infer a first order transition temperature of $T_s \approx 0.06V$.}
  \label{Fig:FirstOrder}
\end{figure}

Below $T \approx 0.1V$ we find two solutions for $m$: the symmetric liquid with $m=0$ and an almost perfect stripe ordered phase with $m \approx 0.5$. Such a jump suggests a first order transition. To further corroborate this claim, we integrate the mean field equation to get the free energy as a function of $m$, see Fig. \ref{Fig:FirstOrder}. The free energy profile clearly shows the presence of a first order transition, with the transition temperature $T_s \approx 0.06V$.

The transition into the stripe phase is so strong that immediately an almost perfect stripe order is created. With a density of states consisting of two delta functions, one expects the resistivity to be simply activated, $\rho_{DC} \sim  4 T \cosh^2 \frac{|V_M|}{4T}$ following Eqn. (\ref{sigmadc}). The first order transition is therefore visible as a jump in the resistivity, see Fig. \ref{Fig:sigmaDC}, which is consistent with experimental results on organic compounds.\cite{Kagawa2013,monceau2007prb,Sato2014PRB}

\section{Supercooled liquid}

The presence of a first order transition, which is confirmed within our classical EDMFT approach, allows for the possibility of a supercooled electron liquid with glassy dynamics. The symmetric $m=0$ solution of the EDMFT equations remains metastable below $T_c$. Upon lowering the temperature, the pseudogap becomes stronger in the supercooled liquid which leads to an exponential suppression of the density of states at the Fermi level, see Fig. \ref{Fig:DOS}. The resulting high resistivity (Fig. \ref{Fig:sigmaDC}) is an indication of kinetic arrest: the motion of electron basically stops and an amorphous glassy state is realized. Such an amorphous state is characterized by short-range correlations, as shown in our Monte Carlo results\cite{Mahmoudian:2014vh} and in X-ray diffuse scattering experiments\cite{Kagawa2013}.

When kinetic arrest itself arises, the system falls out of equilibrium and the manifestly equilibrium theories used here will no longer apply. At these lowest temperatures we expect the self-induced disorder to cause a variable-range hopping type conductivity, yet a more detailed study of this regime remains a challenge for both experimental and theoretical future work.

\section{Conclusion and Outlook}

We have shown, using EDMFT methods, how kinetic arrest of electrons can arise in the presence of long range interactions on a lattice. Our work was motivated by experiments in $\theta$-type organic compounds\cite{Kagawa2013}, though we expect our ideas to be valid in more materials. For example, many metal-insulator transitions are first order which allows supercooling.\cite{Vladbook} In spin systems, recently the pyrochlore 'spin ice' Dy$_2$Ti$_2$O$_7$ has been reinterpreted in terms of a supercooled liquid,\cite{Kassner:2015tm} as well as the suggested spin-orbital liquid Ba$_3$CuSb$_2$O$_9$ \cite{Mila2015}. In all these cases the combination of geometrical frustration and long-range interactions seem to be the key towards understanding supercooled and glassy quantum liquids.

Our results did not require any quenched disorder, which is the standard route for the generation of quantum glasses\cite{Pollak:2013vs}. However, many features seem to be the same: the opening of a soft gap similar to the Coulomb gap\cite{1975JPhC....8L..49E,1976JPhC....9.2021E}, and the Arrhenius-type dynamics.\cite{Mahmoudian:2014vh} It is therefore an interesting question, to be addressed in future research, to what extent self-generated quantum glasses differ from quenched disorder-driven glasses.

\begin{acknowledgements}
L.R. was supported by the
Dutch Science Foundation (NWO) through a Rubicon grant. A.R. and S.F. were supported
by the French National Research Agency through Grant No. ANR-12-JS04-0003-01
SUBRISSYME. V.D. was supported by the NSF grants
DMR-1005751 and DMR-1410132. V. D. would like to thank CPTGA for financing a visit to Grenoble, and KITP at UCSB, where part of the work was performed.
\end{acknowledgements}

\end{document}